# Turbulence energetics in stably stratified geophysical flows: strong and weak mixing regimes

S. S. Zilitinkevich[a-c] *,
T. Elperin[d], N. Kleeorin[d], I. Rogachevskii[d],
I. Esau[c], T. Mauritsen[e], and M. W. Miles[f]

[a] *Division of Atmospheric Sciences and Geophysics, Department of Physics, University of Helsinki, Finland*

[b] *Finnish Meteorological Institute, Helsinki, Finland*

[c] *Nansen Environmental and Remote Sensing Centre / Bjerknes Centre for Climate Research, Bergen, Norway*

[d] *The Pearlstone Centre for Aeronautical Engineering Studies, Department of Mechanical Engineering, Ben-Gurion University of the Negev, Beer-Sheva, Israel*

[e] *Department of Meteorology, Stockholm University, Sweden*

[f] *Environmental System Analysis Research Center, Boulder, USA*

---

* Correspondence address: Finnish Meteorological Institute, Box 503, 00101 Helsinki, FI
E-mail: sergej.zilitinkevich@fmi.fi

**ABSTRACT:** Traditionally, turbulence energetics is characterised by turbulent kinetic energy (TKE) and modelled using solely the TKE budget equation. In stable stratification, TKE is generated by the velocity shear and expended through viscous dissipation and work against buoyancy forces. The effect of stratification is characterised by the ratio of the buoyancy gradient to squared shear, called Richardson number, Ri. It is widely believed that at Ri exceeding a critical value, $Ri_c$, local shear cannot maintain turbulence, and the flow becomes laminar. We revise this concept by extending the energy analysis to turbulent potential and total energies (TPE and TTE = TKE + TPE), consider their budget equations, and conclude that TTE is a conservative parameter maintained by shear in any stratification. Hence there is no "energetics $Ri_c$", in contrast to the hydrodynamic-instability threshold, $Ri_{c\text{-instability}}$, whose typical values vary from 0.25 to 1. We demonstrate that this interval, 0.25<Ri<1, separates two different turbulent regimes: *strong mixing* and *weak mixing* rather than the *turbulent* and the *laminar* regimes, as the classical concept states. This explains persistent occurrence of turbulence in the free atmosphere and deep ocean at Ri>>1, clarify principal difference between turbulent boundary layers and free flows, and provide basis for improving operational turbulence closure models.

## 1. Introduction

In large-scale atmospheric and oceanic flows, variations of the mean velocity, $\bar{\mathbf{u}} = (\bar{u}, \bar{v}, \bar{w})$, density, $\bar{\rho}$, pressure, $\bar{p}$, absolute temperature, $\bar{T}$, and other variables in the vertical (along *z* axis) are usually much larger than in the horizontal (along *x* and *y* axes); and the vertical velocity, $\bar{w}$, is much smaller than horizontal velocities, $\bar{u}$ and $\bar{v}$ (here the overbar and the prime denote mean values and fluctuations; e.g. $u = \bar{u} + u'$). Then, to a good approximation, the mean shear is $\bar{\mathbf{S}} = \mathbf{i}\, \partial \bar{u}/\partial z + \mathbf{j}\, \partial \bar{v}/\partial z$. It causes the shear instability and, at typical geophysical scales, generation of the very-high-Reynolds-number turbulence.

  This process is complicated by the vertical stratification of density. In stable stratification, the mean fluid density $\bar{\rho}$ decreases with increasing height: $\partial \bar{\rho}/\partial z < 0$. Then a fluid element displaced upward (downward) over a distance $\delta z$ differs in density from the ambient fluid by $\rho' = (\partial \bar{\rho}/\partial z)\delta z$ and experiences the downward (upward) acceleration: $(g/\rho_0)\rho' = (g/\rho_0)(\partial \bar{\rho}/\partial z)\delta z$, where $g$ = 9.81 m s$^{-2}$ is the acceleration of gravity and $\rho_0$ is a reference density. In other words, the stable density stratification prevents vertical velocity fluctuations. This effect is the stronger the larger the vertical gradient of the mean buoyancy, $b$, defined as $b \equiv -g\rho/\rho_0$, or its square root $N \equiv (\partial \bar{b}/\partial z)^{1/2}$ called Brunt-Väisälä frequency.

Following Richardson (1920) the relative importance of the counter-effects of shear and stratification are characterised by the dimensionless ratio Ri = $N^2/\mathbf{S}^2$, now called *gradient Richardson number*. Since Richardson's time, a principal question whether velocity shear can or cannot generate turbulence at large Ri has been the focus of attention. It is widely believed, in particular in the meteorological community, that turbulence completely decays when Ri exceeds a critical value, $Ri_c$ (see, e.g. Richardson, 1920; Prandtl, 1930; Taylor, 1931; Chandrasekhar, 1961; Miles, 1961; Monin and Yaglom, 1971; Turner, 1973).

Note that the same symbol ($Ri_c$) and name (critical Richardson number) are applied to the hydrodynamic instability threshold, $Ri_{c\text{-instability}}$, varying from 0.25 to 1 (Taylor, 1931; Miles, 1961; Abarbarnel et al., 1984, 1986; Miles, 1986). As follows from the perturbation analysis, sheared flows are hydrodynamically unstable only at sub-critical Richardson numbers: Ri < $Ri_{c\text{-instability}}$.

At first sight, this leads to the following conclusion: at Ri > $Ri_{c\text{-instability}}$ infinitesimal perturbations are stable − hence the velocity shear cannot generate turbulence. However, this reasoning is inapplicable to finite perturbations: they cause internal gravity waves with inherent orbital motions and local shears, including horizontal shears of vertical velocities, which are not affected by static stability and immediately generate turbulence (Phillips, 1972, 1977). Furthermore, it has been recognised that very short-wave perturbations in sheared flows are dynamically stable even under neutral stratification, so that the stable static stability simply shifts the dynamic instability towards larger wavelengths (Sun, 2006). Hence, perturbation analysis cannot be fully conclusive to answer the question of whether or not the shear can maintain turbulence at large Ri.

Here we emphasise that the "energetics" and the "instability" critical Richardson numbers, $Ri_c$ and $Ri_{c(\text{instability})}$, should not be confused, and limit our analysis to the energetics of turbulence. The overwhelming majority of experiments (see data synthesised in our figures) show general existence of turbulence up to Ri = $10^2$ and do not support the concept of $Ri_c$.

In practical meteorological and oceanographic modelling, this concept, implying no turbulent mixing at Ri > $Ri_c$, is unacceptable. In the free atmosphere, where Ri typically varies from 1 to 10 and often approaches $10^2$, pronounced turbulence has been observed almost continuously at all levels (Lawrence *et al.*, 2004), not to mention that the effective eddy viscosity, $K_M$, and conductivity, $K_H$, are orders of magnitude larger than the molecular ones (Kim and Mahrt, 1992). The same is true for the deep ocean.

To guarantee essential turbulent mixing at large Ri, modern turbulence closures are equipped with Ri-dependencies of the turbulent Prandtl number, $Pr_T \equiv K_M/K_H$, preventing appearance of $Ri_c$, and/or with non-zero background turbulent diffusivities, preventing unrealistic laminarisation.

Some meteorological observations over very cold and smooth surfaces bear witness to a considerable decrease (but never total degeneration) of turbulence in a thin near-surface layer with perceptible wind shears and extremely strong temperature increments (e.g. Smedman *et al.*, 1997). Degeneration of turbulence was occasionally observed in strongly stratified air flows over smooth land surfaces (Monti *et al.*, 2002) and in some lab experiments (Strang and Fernando, 2001). These regimes quite probably correspond to a delayed onset of turbulence due to the absence of pronounced initial perturbations.

## 2. Turbulent energies

Using the state equation and the hydrostatic equation, the density and the buoyancy are expressed in the atmosphere through the potential temperature, $\theta$, and specific humidity, $q$; and in the ocean, through $\theta$ and salinity, $s$. These variables are adiabatic invariants conserved in the vertically displaced portions of fluid, so that the density is also conserved. This allows calculating its fluctuation: $\rho' = (\partial \overline{\rho} / \partial z) \delta z$ and the fluctuation of potential energy per unit mass:

$$\delta E_P = \frac{g}{\rho_0} \int_z^{z+\delta z} \rho' \, dz = \frac{1}{2} \frac{b'^2}{N^2}. \tag{1}$$

For simplicity, we consider the dry, thermally-stratified atmosphere, where the buoyancy, $b$, is expressed through the potential temperature: $b = \beta \theta$ ($\beta = g/T_0$ is the buoyancy parameter, and $T_0$ is a reference value of absolute temperature), whereas the mean-flow equations include only the vertical component, $F_z = \overline{w'\theta'}$, of the potential temperature flux, and the tangential components $\tau_{xz} = \overline{u'w'}$ and $\tau_{yz} = \overline{v'w'}$ of the Reynolds stresses representing the vertical turbulent flux of momentum: $\boldsymbol{\tau} = \mathbf{i}\tau_{xz} + \mathbf{j}\tau_{yz}$.

The familiar budget equations for turbulent kinetic energy (TKE), $E_K = \tfrac{1}{2}\overline{u_i u_i}$, and the mean squared potential temperature fluctuations, $E_\theta = \tfrac{1}{2}\overline{\theta'^2}$, are

$$\frac{DE_K}{Dt} + \frac{\partial \Phi_K}{\partial z} = -\boldsymbol{\tau} \cdot \overline{\mathbf{S}} + \beta F_z - \varepsilon_K, \tag{2}$$

$$\frac{DE_\theta}{Dt} + \frac{\partial \Phi_\theta}{\partial z} = -F_z \frac{\partial \overline{\theta}}{\partial z} - \varepsilon_\theta. \tag{3}$$

Here, $D/Dt = \partial/\partial t + \bar{u}\partial/\partial x + \bar{v}\partial/\partial y$; $t$ is the time; the term $-\boldsymbol{\tau}\cdot\overline{\mathbf{S}}$ describes the TKE production rate; $\Phi_K = \rho_0^{-1}\overline{p'w'} + \tfrac{1}{2}\overline{u_i'u_i'w'}$ and $\Phi_\theta = \tfrac{1}{2}\overline{\theta'^2 w'}$ are the 3rd order vertical turbulent fluxes; $p'$ is the pressure fluctuation; $\varepsilon_K = \nu\overline{(\partial u_i'/\partial x_k')(\partial u_i'/\partial x_k')}$ and $\varepsilon_\theta = -\kappa\overline{\theta'\Delta\theta'}$ are the molecular dissipation rates, $\nu$ is the kinematic viscosity, and $\kappa$ is the temperature conductivity (Tennekes and Lumley, 1972; Kaimal and Fennigan, 1994).

Following Kolmogorov (1941), $\varepsilon_K$ and $\varepsilon_\theta$ are expressed through the turbulent dissipation time scale, $t_T$:

$$\varepsilon_K = E_K(C_K t_T)^{-1}, \quad \varepsilon_\theta = E_\theta(C_P t_T)^{-1}, \tag{4}$$

where $C_K$ and $C_P$ are dimensionless constants of order unity; and $t_T$ can be expressed through the turbulent length scale $l = E_K^{1/2} t_T$.

In view of Eq. (1), $\tfrac{1}{2}(\beta\theta')^2 N^{-2} = \tfrac{1}{2}b'^2 N^{-2}$ is nothing but the fluctuation of potential energy, so that the mean turbulent potential energy (TPE) is defined as $E_P = \tfrac{1}{2}(\beta/N)^2\overline{\theta'^2}$. Then multiplying Eq. (3) by $(\beta/N)^2$ and assuming that $N^2 = \beta\partial\bar\theta/\partial z$ changes only slowly in space and time, gives the following TPE budget equation:

$$\frac{DE_P}{Dt} + \frac{\partial\Phi_P}{\partial z} = -\beta F_z - \varepsilon_P, \tag{5}$$

where $\varepsilon_P = (\beta/N)^2\varepsilon_\theta$ and $\Phi_P = (\beta/N)^2\Phi_\theta$ are the dissipation rate and the vertical turbulent flux of TPE.

In a sense, TPE is analogous to the available potential energy (APE) introduced by Lorenz (1955, 1967): both APE and TPE are proportional to the squared perturbation of potential temperature or density (in contrast to the seemingly natural idea of the linear dependence of any kind of potential energy on density). The principal difference between these two concepts is that APE is an integral property of the entire flow-domain (e.g. of the atmosphere as a whole), whereas TPE is determined in each point of turbulent flow.

The term $\beta F_z$ appears in Eqs. (2) and (5) with opposite signs and describes the energy exchange between TKE and TPE. It annihilates in the budget equation for the total turbulent energy (TTE), $E = E_K + E_P$, which has the form of a conservation equation:

$$\frac{DE}{Dt} + \frac{\partial\Phi_E}{\partial z} = -\boldsymbol{\tau}\cdot\overline{\mathbf{S}} - \varepsilon_E, \tag{6}$$

where $\varepsilon_E = \varepsilon_K + \varepsilon_P$ and $\Phi_E = \Phi_K + \Phi_P$ are the dissipation rate and the vertical turbulent flux of TTE.

The left-hand sides of Eqs. (2), (3), (5) and (6) are neither productive nor dissipative and describe the energy transports. In the equilibrium (homogeneous and stationary) regime they turn into zero, so that the TTE budget Equation (6) simplifies to $\varepsilon_E = -\mathbf{\tau} \cdot \mathbf{\bar{S}} > 0$, which implies generation of TTE in any stratification and thus argues against any finite value of the energetics critical Richardson number [cf. Eq. (11) below].

In view of Eqs. (2)–(6), maintaining turbulence at large Ri can be explained as follows. Suppose that the buoyancy flux, $\beta F_z$, becomes so large that TKE considerably decreases. According to Eq. (6), TTE conserves, so that TPE increases and fluctuations of buoyancy strengthen. In other words, fluid elements acquire stronger accelerations and speed up toward their "equilibrium level", which causes re-establishing of TKE, and decreasing of TPE. In its turn, too large TKE causes stronger displacements of fluid elements, hence stronger buoyancy fluctuations and therefore increasing of TPE. Such oscillations are typical of intermittent turbulence.

The TPE fraction, $E_P/E$, is negligible in neutral stratification and increases with strengthening static stability (increasing Ri). Generally speaking, the dependence of $E_P/E$ on Ri is not universal. However, in the equilibrium turbulence regime, when the left-hand sides of the energy budget equations become zero, Eqs. (4)-(6) yield a simple dependence of $E_P/E$ on the so-called flux Richardson number, $\mathrm{Ri}_f = \beta F_z (\mathbf{\tau} \cdot \mathbf{\bar{S}})^{-1}$:

$$\frac{E_P}{E} = \frac{(C_P/C_K)\mathrm{Ri}_f}{1 + (C_P/C_K - 1)\mathrm{Ri}_f}; \tag{7}$$

the budget equations for the turbulent fluxes simplify to the familiar down-gradient formulations (Monin and Yaglom, 1971; Tennekes and Lumley, 1972):

$$\mathbf{\tau} = -K_M \mathbf{\bar{S}}, \quad \beta F_z = -K_H N^2; \tag{8}$$

and the flux Richardson number becomes:

$$\mathrm{Ri}_f = \mathrm{Ri}/\mathrm{Pr}_T. \tag{9}$$

Furthermore, the turbulent Prandtl number, $\mathrm{Pr}_T \equiv K_M/K_H$, and in view of Eqs. (7)-(8) – the TPE fraction, $E_P/E$, become universal functions of Ri [see Zilitinkevich *et al.* (2007)].

Until present quantitative analyses of the turbulence energetics were basically limited to the TKE budget. Only recently has the Eq. (3) for the squared potential-temperature fluctuations been treated in terms of TPE (e.g. Holloway, 1986; Dalaudier and Sidi, 1987; Hunt *et al.*, 1988; Canuto and Minotti, 1993; Schumann and Gerz, 1995; Hanazaki and Hunt, 1996; Keller and van Atta, 2000; Canuto *et al.,* 2001; Stretch *et al.*, 2001; Cheng et al., 2002; Luyten et al., 2002; Jin *et al.*, 2003; Hanazaki and Hunt, 2004; Rehmann and Hwang, 2005; Umlauf, 2005). The budget equations for all three energies, TKE, TPE and TTE, were considered by Canuto and Minotti (1993), Canuto *et al.* (2008), Elperin *et al.* (2002) and Zilitinkevich *et al.* (2007). Zilitinkevich (2002) employed the pair of budget equations – for TKE and TPE – to derive a non-local closure model, which allowed explaining the distant effect of the free-flow stability on the surface-layer turbulence.

Clearly, turbulent flows, as any other mechanical systems, are not fully characterised by their kinetic energy. It is not surprising that the traditional approach based on the solely considered TKE budget could be misleading. The concept of the energetics critical gradient Richardson number is one example. In due time it was deduced from Eq. (2) as follows: in very strong static stability (at large Ri) the negative buoyancy flux, $\beta F_z$, passes a threshold, after which the TKE production, $-\boldsymbol{\tau}\cdot\mathbf{S}$, becomes insufficient to compensate the TKE losses, $-\beta F_z + \varepsilon_K$, so that the turbulence could only decay (Prandtl, 1930; Chandrasekhar, 1961; Monin and Yaglom, 1971).

However, the steady-state TKE budget equation, $-\boldsymbol{\tau}\cdot\mathbf{S} = -\beta F_z + E_K(C_K t_T)^{-1}$, is not closed. The above reasoning says only that the ratio of the TKE consumption to its production, $\text{Ri}_f = -\beta F_z/(-\boldsymbol{\tau}\cdot\mathbf{S})$ called flux Richardson number, cannot exceed unity. But $\text{Ri}_f$ is an internal turbulent parameter ($\boldsymbol{\tau}$ and $F_z$ depend on each other), which is why the restriction $\text{Ri}_f < 1$ says nothing about maintenance or degeneration of turbulence at large Ri. To proceed further, the traditional approach employs Eqs. (8)-(9) and assumes that the turbulent Prandtl number, $\text{Pr}_T$, is either constant or limited to a finite maximal value, $\text{Pr}_{T-max}$. If so, it would indeed follow from the TKE budget equation that the equilibrium turbulence exists only at Ri smaller than some critical value $\text{Ri}_c < \text{Pr}_{T-max}$.

The fallacy in this conclusion is that neither theory nor experiments confirm the existence of any upper limit for $\text{Pr}_T$. On the contrary, the presence of turbulence at very large Ri has been disclosed in numerous experiments and numerical simulations, in particular those summarised in Figures 1-4 below. Furthermore, Figure 3 (and prior empirical evidence) clearly shows unlimited increase of $\text{Pr}_T$ with increasing Ri.

## 3. Strong- and weak-mixing regimes

Below we consider empirical data on TKE, TPE and vertical turbulent fluxes of momentum and potential temperature for different stratification regimes from neutral (Ri = 0) to very stable (Ri >> 1).

Figure 1 shows $E_P/E$ as dependent on Ri after recent atmospheric experiments (Uttal *et al.*, 2002), lab experiments (Ohya, 2001), and our large eddy simulations (LES) using NERSC code (Esau, 2004). In the experiments, very large Ri are observed above the turbulent boundary layers, in the strongly heterogeneous "capping" temperature inversions, where considerable amounts of TKE and TPE are transported from the boundary-layer interior, so that the energetics of turbulence is not fully controlled by local factors, and $E_P/E$ depends not only on Ri but to a large extent on the initial and boundary conditions.

In this context quite useful are LES representing more homogeneous regimes, where the basic features of turbulence are more closely linked with the focal factors. In particular, LES data in Figure 1 show a well-pronounced monotonic dependence: the ratio $E_P/E$ sharply increases with increasing Ri in the interval 0 < Ri < 1 and then levels off approaching the limiting value: $E_P/E \approx 0.25$.

Figure 2 shows the Ri-dependences of the normalised turbulent fluxes of momentum, $\tau/E_K$ (where $\tau = |\boldsymbol{\tau}|$), and heat, $-F_z/(E_K E_\theta)^{1/2}$ – once again better pronounced in LES data.

Figure 3 shows the Ri-dependence of the turbulent Prandtl number, $\Pr_T$, after recent data from literature (Monti et al., 2002; Strang and Fernando, 2001; Ohya, 2001; Bange and Roth, 1999; Rehmann and Koseff, 2004; Stretch et al., 2001; Mauritsen and Svensson, 2007) and our LES. Data for very large Ri follow the linear law: $\Pr_T \approx 5\text{Ri}$; data for very small Ri approach well-known neutral-stability limit: $\Pr_T \approx 0.8$ (Churchill, 2002); and data for any Ri are roughly approximated by the interpolation formula:

$$\Pr_T \approx 0.8 + 5\text{Ri}. \tag{10}$$

The latter implies that the flux Richardson number, $\text{Ri}_f = \text{Ri}/\Pr_T$, monotonically increases with increasing Ri and approaches $\text{Ri}_f = \text{Ri}_f^\infty \approx 0.2$ at Ri >> 1.

Generally speaking Figure 3 could suffer from the artificial self-correlation between $\Pr_T$ determined as $(\tau N^2)/(\beta F_z S)$ and $\text{Ri} = (N/S)^2$. However, the small-Ri interval (Ri < $10^{-1}$) is obviously free from this drawback, which lends credence to the entire figure. Moreover, the linear Ri-dependence of the turbulent Prandtl number represents the only physically consistent very-large-Ri asymptote. The higher and the lower power laws are both unacceptable: $\Pr_T \sim \text{Ri}^{1+\varepsilon}$ substituted into Eq. (9) leads to the physically senseless decrease of $Ri_f$ with increasing Ri; whereas $Pr_T \sim Ri^{1-\varepsilon}$ leads to the limitless increase of $Ri_f$ up to $Ri_f > 1$, which contradicts Eq. (2).

Using empirical very-large-Ri limits disclosed in Figures 1 and 3, namely $E_P/E \approx 0.3$ and $Ri_f \approx 0.2$, Eq. (7) allows estimating the ratio of the dissipation constants in Eq. (4): $C_K/C_P \approx 0.6$. Then, using empirical large-Ri limits: $E_P/E \approx 0.25$ and $E_K/E = (E - E_P)/E \approx 0.7$ after Figure 1 and $Ri_f \approx 0.2$ after Figure 3, Eq. (4) for the dissipation rates yields $\varepsilon_E \approx 0.7\, C_K^{1/2} E^{3/2} l^{-1}$. Then using the very-large-Ri limit: $\tau/E_K \approx 0.1$ after Figure 2, the equilibrium TTE budget equation, $\varepsilon_E = -\boldsymbol{\tau} \cdot \overline{\mathbf{S}}$, yields the asymptotic formula:

$$E \approx 0.02(C_K S l)^2 > 0 \quad \text{at} \ \ Ri \gg 1. \qquad (11)$$

Equation (11) determines essentially positive TTE in any stationary, homogeneous sheared flow and confirms our argumentation against the energetics critical Richardson number.

LES data in Figure 2 reveal that $\tau/E_K$ as well as $-F_z/(E_K E_\theta)^{1/2}$ turn into constants in the two alternative regimes: near-neutral and very stable with the sharp transition in the narrow interval of Ri around $Ri \approx 0.2$–$0.3$ (cf. Mahrt et al., 1998). The same kind of transition between $E_P/E = 0$ and $E_P/E \approx 0.3$ is recognisable in Figure 1. Moreover, as seen from Figure 3, the two asymptotes: $Pr_T \approx 0.8$ for $Ri \ll 1$ and $Pr_T \approx 5\,Ri$ for $Ri \gg 1$ match at $Ri \sim 0.25$. Coincidence of this value with the classical hydrodynamic instability threshold is eye-catching. However, as our figures prove, this threshold by no means separates the turbulent and the laminar regimes, as the classical concept stated, but the two essentially different turbulent regimes:

- (Ri < 0.1) *strong mixing* capable of very efficiently transporting both momentum: $\tau/E_K \approx 0.3$ and heat: $-F_z/(E_K E_\theta)^{1/2} \approx 0.4$;
- (Ri > 1) *weak mixing* quite capable of transporting momentum: $\tau/E_K \to$ constant $\approx 0.1$; but rather inefficient in transporting heat: $-F_z/(E_K E_\theta)^{1/2}$ drops to ~0.04 at Ri = 50 and, as follows from the asymptotic analysis of Eq. (3), tends to zero as $Ri^{-1/2}$ at $Ri \gg 1$ (in accordance with Figure 3).

It is conceivable that the weak turbulence regime is most probably dominated by internal waves, which efficiently transport momentum but do not transport heat (see, e.g. Nappo, 2002). For large Richardson numbers, the source of turbulence can be either internal gravity waves or so-called pancake vortices (see Lilly, 1983). Thus the terms "strong" and "weak" acquire concrete physical sense: strong turbulence is fully chaotic and vortical, whereas weak turbulence is wave dominated and presumably intermittent.

Among practically important applications of turbulence closures suitable for very stable stratification we mention the deep-ocean downward heat flux known to be a controlling factor of the rate of global warming (Hansen *et al*., 1985) and optical turbulence in the free atmosphere essential for astronomical observations (Lawrence *et al*., 2004).

The above analyses disprove the concept of the "energetics" critical Richardson number in its classical sense. Experimental, LES and DNS data summarised in our figures, and other evidences from modern literature (e.g. Zilitinkevich *et al*., 2007; Mauritsen *et al*., 2007; Canuto *et al*., 2008; Galperin *et al*., 2007) demonstrate general existence of turbulence at very large Ri, up to Ri > $10^2$ exceeding its commonly accepted critical values by more than two orders of magnitude.

What is factually observed is a threshold interval of Richardson numbers, 0.1 <Ri <1, separating two regimes of essentially different nature but both turbulent. The laminar regime could take place at very large Ri in the absence of pronounced initial perturbations, most probably due to the delayed onset of turbulence.

The concept of the two principally different turbulent regimes sheds light upon many uncertain problems. In particular, it allows refining the definition of the stably stratified atmospheric boundary layer (ABL) as the *strong-mixing* stable layer, in contrast to the also stable but *weak-mixing* free atmosphere. Because these two turbulent regimes are characterised by the small and the large Ri, respectively, it is natural to expect that the ABL outer boundary, $z = h$, should fall into the threshold interval: $0.1 < \text{Ri} < 1$.

Figure 4 confirms this conclusion. It shows Ri as dependent on the dimensionless height *z/L*, where $L = \tau^{3/2}(-\beta F_z)^{-1}$ is the Monin-Obukhov length scale (Monin and Obukhov, 1954) widely used in boundary-layer meteorology. Red and pink points show our LES data: red, for the ABL interior ($z < h$); pink, for the free atmosphere ($z > h$), where $h$ is determined as the height at which $\tau$ diminishes to 5% of its surface value [atmospheric data (Uttal *et al*., 2002, blue points) do not give such opportunity].

## 4. Concluding remarks

Our analyses demonstrate that the TKE budget equation, used solely in many theoretical analyses and applications, is not sufficient to characterise the energy transformations in stratified flows. Equally important are the budget equations for TPE and TTE.

The latter, in contrast to the kinetic energy, represents an invariant conserved in the absence of the shear and dissipation. Its budget equation provides the key argument against the energetics critical Richardson number and opens new ways towards advancing turbulence closures and computational tools for geophysical fluid mechanics.

Data analyses in Section 3 represent illustrations rather than validation of our basic conclusions. Further experimental and numerical-simulation studies of similar kind are needed.

## Acknowledgements

This work has been supported by EU FP7 Project MEGAPOLI No. 212520; EU Project TEMPUS JEP 26005; Academy of Finland Project IS4FIRES; ARO Project W911NF-05-1-0055, Norwegian Research Council projects NORCLIM 178246 and POCAHONTAS 178345/S30; German-Israeli Project Cooperation DIP; Israel Science Foundation; and Carl-Gustaf Rossby International Meteorological Institute in Stockholm. We acknowledge discussions with Vittorio Canuto (USA), Eero Holopainen (Finland), Ola M. Johannessen (Norway), Victor L'vov (Israel) and Arkady Tsinober (UK).

# Figure captions

Figure 1. The ratio of the potential to total turbulent energies, $E_P/E$, versus the gradient Richardson number, Ri. Blue points and curve – meteorological field campaign SHEBA (Uttal *et al*., 2002); green – lab experiments (Ohya, 2001); red/pink – new large-eddy simulations (LES) using NERSC code (Esau, 2004). Vertical error bars show one standard deviation above and below the averaged value within the bin; horizontal bars show the width of the bins.

Figure 2. Normalised turbulent fluxes of momentum and heat, $\tau/E_K$ (a) and $F_z/(E_K E_\theta)^{1/2}$ (b), versus Ri after the same data as in Figure 1.

Figure 3. Turbulent Prandtl number $\Pr_T = K_M/K_H$ versus Ri. Blue points and curve – meteorological campaigns SHEBA (Uttal *et al*., 2002, mostly for Ri < 1) and CASES-99 (Poulos *et al*., 2002, for 0.1 < Ri <100); green – laboratory sheared flow (Ohya, 2001); red – new LES using NERSC code (Esau 2004); grey – direct numerical simulations (DNS) with 32 (lightest), 64 (darker) and 128 (darkest) nodes, respectively (Stretch *et al*., 2001). Numbers show data from literature: **1** – nocturnal atmospheric boundary layer (Bange and Roth, 1999); **2** – sediment-loaded flow (COSINUS 2000); **3** – laboratory turbulence (Polyakov, 1989); **4** – laboratory grid-generated turbulence (Rehmann and Koseff, 2004); **5** – laboratory sheared flow (Strang and Fernando, 2001); **6** – atmospheric slope flow (Monti *et al*., 2002). The dashed curve: $\Pr_T = 0.8 + 5\mathrm{Ri}$ is composed of the two asymptotes: already known: $\Pr_T = 0.8$ at Ri < 0.1, and obtained from this figure: $\Pr_T = 5\mathrm{Ri}$ at Ri >1. Red, green and blue curves show bin-averaged data for the corresponding data sources. Horizontal bars show the width of bins. Vertical bars show one standard deviation above and below the averaged value within the bin. The thin line: $\mathrm{Ri}/\Pr_T = \mathrm{Ri}_f = 1$ separate the "principally impossible area" ($\mathrm{Ri}_f$ cannot exceed unity in the steady state).

Figure 4. The gradient Richardson number within and above the stable ABL: Ri versus *z/L*, where $L = \tau^{3/2}(-\beta F_z)^{-1}$ is the Monin–Obukhov length scale. Red points (for *z* < *h*) and pink points (for *z* > *h*) show LES data [NERSC code, Esau (2004)]; blue points show atmospheric data (Uttal *et al*., 2002).

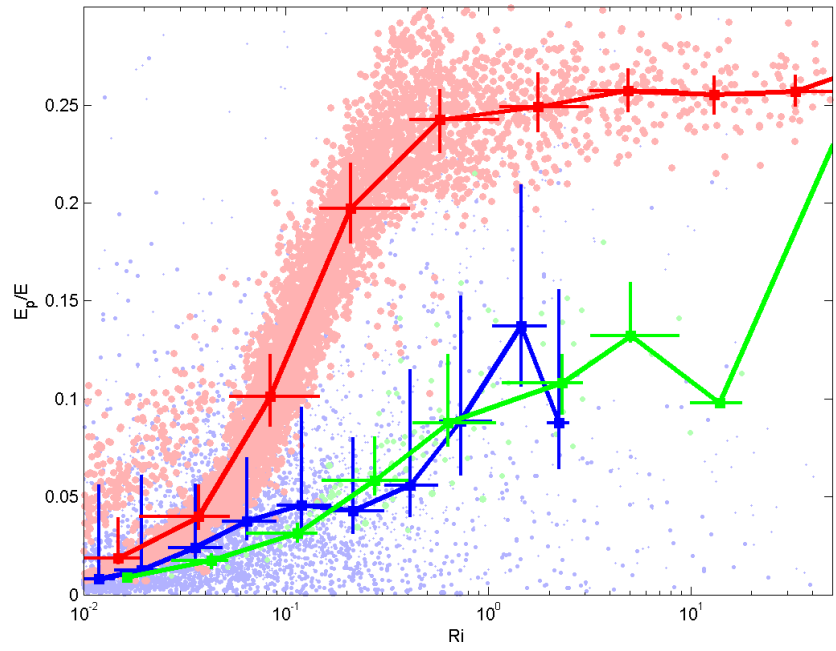

**Figure 1.**

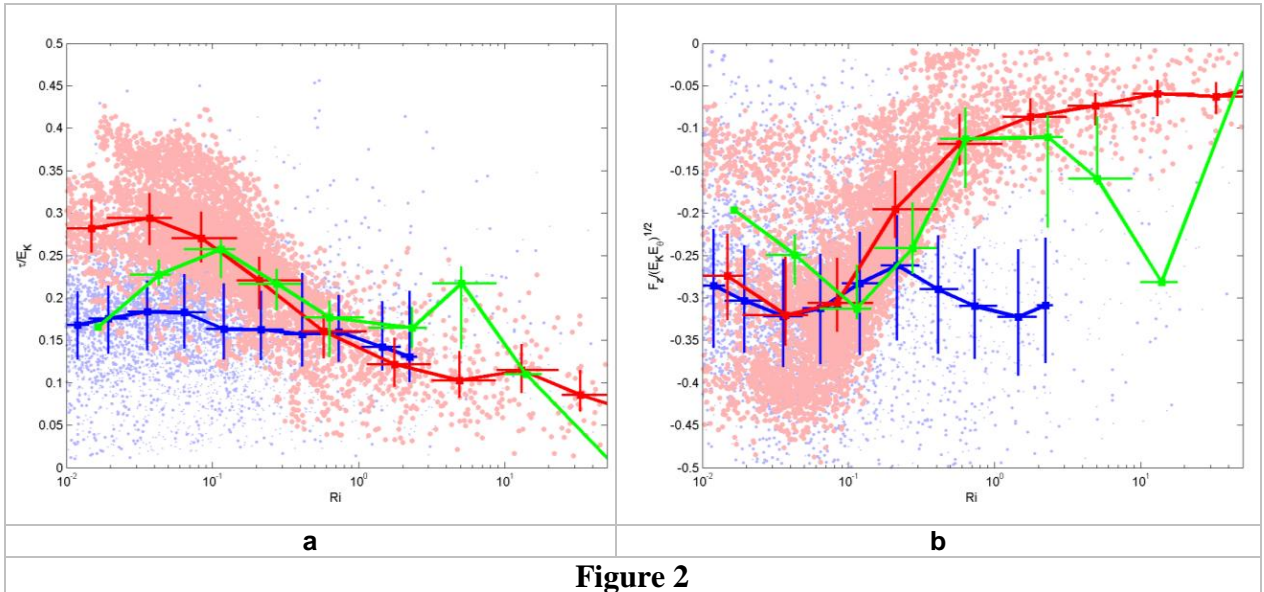

**Figure 2**

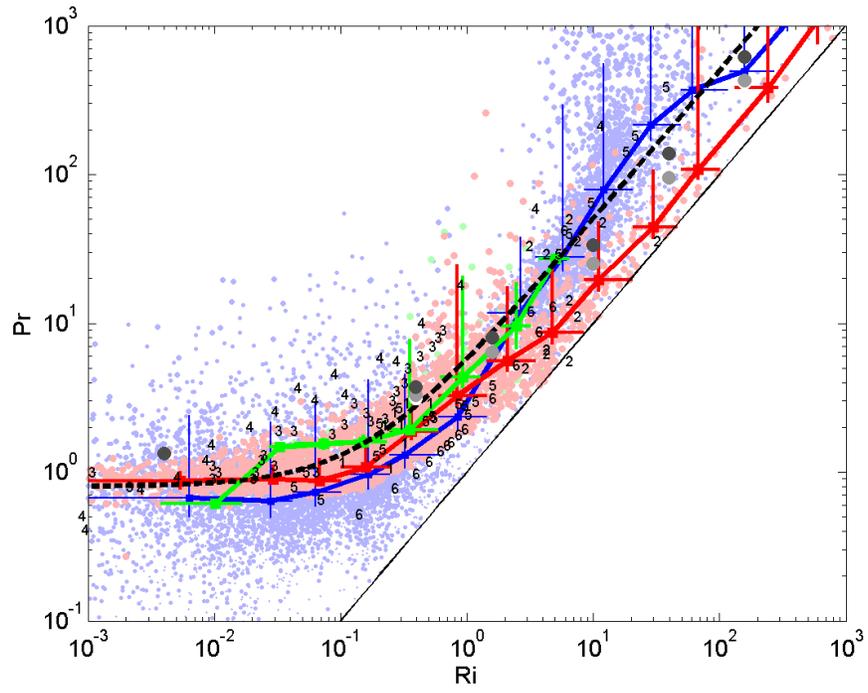

**Figure 3.**

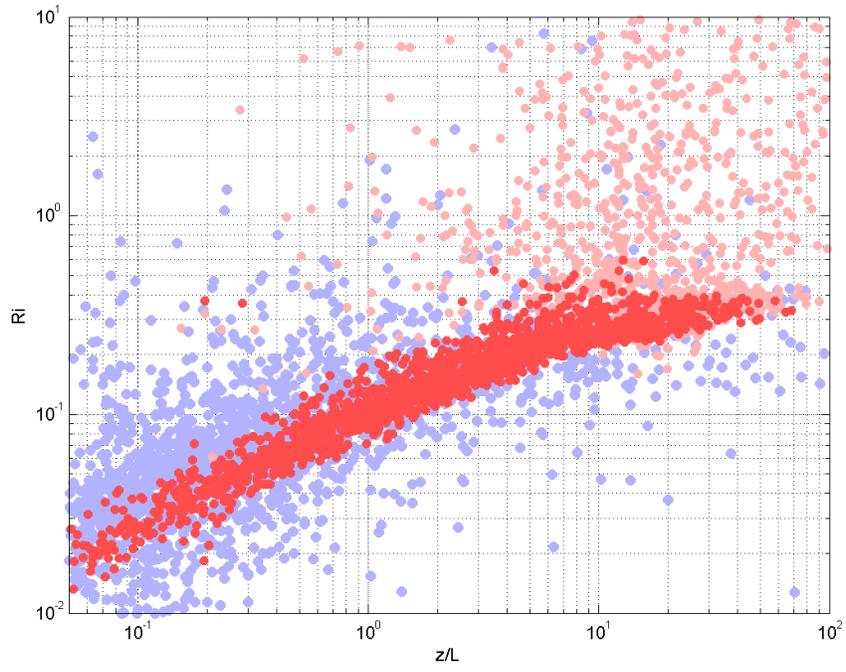

**Figure 4.**